\documentclass[aps,prc,preprintnumbers,superscriptaddress,twocolumn]{revtex4}

\usepackage{latexsym}
\usepackage{amssymb}
\usepackage{amsmath}
\usepackage{graphicx}
\usepackage{bm}
\usepackage{epsf}
\usepackage{epstopdf}

\newcommand{\MeV}{\mathrm{MeV}}

\usepackage[usenames]{color}
\usepackage[normalem]{ulem}  %

\newcommand{\comment}[1]{}

\renewcommand\sout{\bgroup \color{red} \ULdepth=-.5ex \ULset}

\begin{document}

\title{Symmetry energy effect on the secondary component of GW190814 as a neutron star}
\author{Xuhao Wu} ~\email{x.h.wu@pku.edu.cn}
\affiliation{School of Physics, Peking University, Beijing 100871, China}
\affiliation{Kavli Institute for Astronomy and Astrophysics, Peking University, Beijing 100871, China}
\author{Shishao Bao} ~\email{bao_shishao@163.com}
\affiliation{School of Physics and Information Engineering, Shanxi Normal University, Linfen 041004, China}
\author{Hong Shen}~\email{shennankai@gmail.com}
\affiliation{School of Physics, Nankai University, Tianjin 300071, China}
\author{Renxin Xu}~\email{r.x.xu@pku.edu.cn}
\affiliation{School of Physics, Peking University, Beijing 100871, China}
\affiliation{Kavli Institute for Astronomy and Astrophysics, Peking University, Beijing 100871, China}

\begin{abstract}
The secondary component of GW190814 with a mass of 2.50-2.67 $M_{\odot}$ 
may be the lightest black hole or the heaviest neutron star 
ever observed in a binary compact object system. 
To explore the possible equation of state (EOS), which can support such massive neutron star, 
we apply the relativistic mean-field model with 
a density-dependent isovector coupling constant to describe the neutron-star matter. 
The acceptable EOS should satisfy some constraints:  
the EOS model can provide a satisfactory description of the nuclei;
the maximum mass $M_\textrm{TOV}$ is above 2.6 $M_{\odot}$; 
the tidal deformability of a canonical 1.4 $M_{\odot}$ neutron star $\Lambda_{1.4}$ 
should lie in the constrained range from GW170817.
In this paper, we find that the nuclear symmetry energy 
and its density dependence play a crucial role in determining the EOS of neutron-star matter. 
The constraints from the mass of 2.6 $M_{\odot}$ 
and the tidal deformability $\Lambda_{1.4}=616_{-158}^{+273}$ 
(based on the assumption that GW190814 is a neutron star-black hole binary) 
can be satisfied as the slope of symmetry energy $L \leq 50$ MeV. 
Even including the constraint of $\Lambda_{1.4}=190_{-120}^{+390}$ from GW170817 
which suppresses the EOS stiffness at low density,
the possibility that  the secondary component of GW190814 is a massive neutron star
cannot be excluded in this study. 
\end{abstract}

\maketitle

\section{Introduction}
\label{sec:1}

The first direct detection of gravitational waves from GW150914~\cite{Abbott2016}, 
the merger of a pair of black holes, launched a new era of gravitational-wave astronomy.
Two years later, the binary neutron star merger event GW170817~\cite{Abbott2017}, 
which was observed by the LIGO and Virgo detectors, 
produced a detectable electromagnetic signal 
and marked a significant breakthrough for multi-messenger astronomy.
After that, another 
binary neutron star merger event GW190425 
was detected in April of 2019~\cite{Abbott2020a}. 
Several months after GW190425, in the third observing run of 
Advanced LIGO and Advanced Virgo~\cite{Abbott2020b}, 
a gravitational wave signal GW190814 was discovered 
from a compact binary coalescence involving a 22.2-24.3~$M_\odot$ black hole 
and a compact object of 2.50-2.67~$M_\odot$. 
Since no measurable tidal signature was detected from the gravitational waveform 
and no electromagnetic counterpart of GW190814 has been confirmed,
the secondary component of GW190814 could be either the heaviest neutron star 
or the lightest black hole ever discovered~\cite{Abbott2020b}. 
Although the highest mass measured of radio pulsars is around 2.14~$M_\odot$~\cite{Cromartie2020}, 
the EOS should be very stiff at high density if neutron-star maximum mass $M_\textrm{TOV} > 2.3~M_\odot$~\cite{Wu2020}.

It seems difficult to determine the nature of GW190814 by further analyzing the data, 
that many theoretical studies have been devoted to exploring 
various possibilities of the secondary component of GW190814. 
A Markov Chain Monte Carlo approach was raised by Godzieba et al.~\cite{Godzieba2020} 
to generate phenomenological EOSs that could meet the astronomical constraints 
and support the GW190814 is a neutron star-black hole system. 
Fattoyev et al.~\cite{Fattoyev2020} proposed the BigApple parameter set 
in covariant energy density functional theory, 
which predicts the maximum neutron-star mass of 2.6~$M_\odot$ and can 
reproduce the observables of finite nuclei and NICER. 
The appearance of  deconfined QCD matter in the neutron star 
may support a massive enough neutron star $M_\textrm{TOV} > 2.5$~$M_\odot$~\cite{Tan2020}. 
This view is also indirectly supported by the quarkyonic matter influence on the EOS~\cite{McLerran2019}.
 
To assess the nature of the secondary component of GW190814, 
the EOS at high density plays an essential role.
It is well known that the EOS is crucial in determining 
the mass-radius relation of neutron stars.
Due to the intractability of treating the interactions 
in nuclear many-body systems,
phenomenological mean-field models have been 
widely used for describing the EOS of neutron stars.
%
In this work, we use the relativistic mean-field (RMF) model 
to describe the neutron-star matter, 
in which nucleons interact via the exchange of various mesons.
Among these mesons, the scalar meson $\sigma$ describes attraction between baryons, 
the isoscalar-vector meson $\omega$ describes repulsion, 
and the isovector-vector meson $\rho$ is 
included to account for the isospin asymmetry.
The pseudoscalar meson $\pi$, which has the primary role 
in long-range baryon-baryon interaction, 
vanishes in the RMF approximation due to its odd parity. 
The RMF model parameters are typically determined 
by fitting the experimental properties of finite nuclei 
or the empirical saturation properties at the nuclear saturation density $n_0$, i.e., 
the binding energy per nucleon ($E/A$), the nuclear incompressibility ($K$), 
the symmetry energy ($S$) and the effective mass ($m^*/m$). 
The symmetry energy slope ($L$), which has an apparent effect on the neutron-star radius 
and the neutron-skin thickness of neutron-rich nuclei~\cite{Centelles2009,Fattoyev2018,Hu2020}, 
is another important parameter that has been extensively studied
in recent years~\cite{Steiner2012,Lattimer2013,Tews2017,Oertel2017}.
However, its value is still very uncertain 
and cannot be well constrained from current observations.
In order to explore the influence of symmetry energy slope, 
one may introduce density-dependent isovector couplings 
or add $\omega$-$\rho$ coupling term~\cite{Bao2014}. 
The two approaches are basically equivalent~\cite{Drago2014}.
In this paper, we will use the RMF model 
with a density-dependent isovector coupling constant 
(referred to as the RMFL model following Ref.~\cite{Spinella2017}).
Through this way, the symmetry energy slope $L$ can be tailored 
by adjusting an additional coefficient without affecting the other saturation properties, 
and remaining all parameters unchanged.

Over the past decade, several massive neutron stars have been discovered,
PSR J1614-2230~\cite{Demorest2010,Fonseca2016,Arzoumanian2018}, 
PSR J0348+0432~\cite{Antoniadis2013}, PSR J0740+6620~\cite{Cromartie2020}, 
and PSR J2215+5135~\cite{Linares2018}, 
which imposed a lower bound to the maximum mass of neutron stars ($M_\textrm{TOV}>2~M_\odot$). 
Moreover, the binary neutron star merger event GW170817 
provides new constraints on the tidal deformability of a canonical
1.4 $M_\odot$ neutron star,
$\Lambda_{1.4} \le 800$ reported in the discovery paper~\cite{Abbott2017}
and the updated value of  $\Lambda_{1.4}=190_{-120}^{+390}$ in Ref.~\cite{Abbott2018}. 
The radius of a 1.4 $M_\odot$ neutron star $R_{1.4}$ 
could be constrained to be $R_{1.4} \le 13.6$~km 
with different models~\cite{Annala2018, Most2018}. 
Besides, based on NASA's Neutron Star Interior Composition Explorer (NICER)
data set, 
an estimation between mass-radius relation becomes possible 
by using X-ray pulse-profile modeling~\cite{Raaijmakers2019}. 
Through this way, the mass and radius of PSR J0030+0451 were 
reported as ($1.44^{+0.15}_{-0.14}$~$M_\odot$, $13.02^{+1.24}_{-1.06}$~km)~\cite{Miller2019} 
or ($1.34^{+0.15}_{-0.16}$~$M_\odot$, $12.71^{+1.14}_{-1.19}$~km)~\cite{Riley2019} 
respectively by different groups. 
Among these observation properties, neutron star mass and tidal deformability 
are dynamical and model-independent measurements that provide strong constraints, 
while the measurement of the radius may have different results with various models.
It is believed that the binary neutron star merger event GW170817 may eventually
become a black hole. Based on this assumption, upper bounds on
$M_\textrm{TOV}$ supported by the EOS are placed to be $M_\textrm{TOV}$ $\sim$ 2.3 ~$M_\odot$~\cite{Margalit2017,Ruiz2018,Rezzolla2018,Shibata2019,Abbott2020c}. 
But without this assumption on the remnant, 
larger $M_\textrm{TOV}$ could be achieved~\cite{Abbott2020c}.
Besides, $M_\textrm{TOV} >2.3~M_\odot$ can not be ruled out 
from the views of microphysics and astrophysics.

In this paper, we aim to use the RMFL model with NL3 parametrization 
to generate a series of density-dependent isovector coupling parameter sets 
(referred to as NL3L) with different symmetry energy slope $L$. 
The original NL3 parametrization was proposed by fitting the total binding energies, 
the charge radii, and the available neutron radii of nuclei~\cite{Lalazissis1997}.
The generated NL3L parameter sets are expected to achieve 
similar ground-state properties of finite nuclei as NL3.
Using NL3L parameter sets, we examine the role of 
symmetry energy slope $L$ in neutron-star matter EOS and properties of neutron stars. 
It is believed that a smaller $L$ corresponds to a softer EOS 
at the density range from $n_0$ to $\sim 0.3 ~\textrm{fm}^{-3}$,
which mainly affects the radius and tidal deformability of neutron star with $M < 1.5~M_\odot$.


In general, the presence of exotic degrees of freedom,
like hyperons and quarks, tends to soften the EOS at high densities 
and reduce the maximum mass of neutron stars~\cite{Ishizuka2008,Wu2018,Wu2017}.
However, this effect can be suppressed by considering the additional repulsion for hyperons (or quarks).
Furthermore, the crossover hadron-quark phase transition may provide a stiffer quarkyonic core in neutron star. 
For simplicity, we do not include non-nucleonic degrees of freedom in the present work. 
We aim to investigate the symmetry energy effect on the properties of
massive neutron stars, and explore the possibility of the secondary component
of GW190814 as a neutron star.

This article is organized as follows.
In Sec.~\ref{sec:rmfl}, we briefly introduce
the RMFL model for neutron-star matter.
In Sec. \ref{sec:results}, we show
the numerical results of neutron-star properties,
and discuss the constraints from astronomical observations.
Section \ref{sec:summary} is devoted to a summary.

\section{RMF with a Density-Dependent Isovector Coupling Constant}
\label{sec:rmfl}
\begin{table*}[htp]
	\caption{Parameters in the NL3 model.
		The masses are given in MeV.}
	\begin{center}
		\setlength{\tabcolsep}{2.6mm}{
			\begin{tabular}{lcccccccccccc}
				\hline\hline
				Model   &$M$  &$m_{\sigma}$  &$m_\omega$  &$m_\rho$  &$g_\sigma$  &$g_\omega$
				&$g_\rho(n_0)$ &$g_{2}$ (fm$^{-1}$) &$g_{3}$   \\
				\hline
				NL3     &939.000  &508.194  &782.501  &763.000  &10.217  &12.868  &8.948
				&$-$10.431   &-28.885      \\
				\hline\hline
		\end{tabular}}
		\label{tab:nl3}
	\end{center}
\end{table*}
\begin{table}[htp]
	\caption{Nuclear matter saturation properties obtained in the NL3 model. All quantities are given in MeV, except $n_0$, which is given in fm$^{-3}$.}
	\begin{center}
		\setlength{\tabcolsep}{2.4 mm}{
			\begin{tabular}{lcccccccccccc}
				\hline\hline
				Model   &$n_0$  &$E/A$  &$K$  &$S$  &$L$  &$m^*/m$
				\\
				\hline
				NL3     &0.148  &-16.24  &272.3  &37.4  &118.5  &0.594
				\\
				\hline\hline
		\end{tabular}}
		\label{tab:sat}
	\end{center}
\end{table}
We adopt the RMFL model to describe the neutron-star matter, 
in which the isovector coupling constant is taken to be
density dependent as in the density-dependent RMF (DDRMF) approach.
We use the Lagrangian given as
\begin{eqnarray}
\label{eq:LRMF}
\mathcal{L}_{\rm{RMFL}} & = & \sum_{i=p,n}\bar{\psi}_i
\bigg \{i\gamma_{\mu}\partial^{\mu}-\left(M+g_{\sigma}\sigma\right)
\notag\\
&&-\gamma_{\mu} \left[g_{\omega}\omega^{\mu} +\frac{g_{\rho}}{2}\tau_a\rho^{a\mu}
\right]\bigg \}\psi_i  \notag \\
&& +\frac{1}{2}\partial_{\mu}\sigma\partial^{\mu}\sigma -\frac{1}{2}%
m^2_{\sigma}\sigma^2-\frac{1}{3}g_{2}\sigma^{3} -\frac{1}{4}g_{3}\sigma^{4}
\notag \\
&& -\frac{1}{4}W_{\mu\nu}W^{\mu\nu} +\frac{1}{2}m^2_{\omega}\omega_{\mu}%
\omega^{\mu}   \notag
\\
&& -\frac{1}{4}R^a_{\mu\nu}R^{a\mu\nu} +\frac{1}{2}m^2_{\rho}\rho^a_{\mu}%
\rho^{a\mu}
\notag\\
&& +\sum_{l=e,\mu}\bar{\psi}_{l}
  \left( i\gamma_{\mu }\partial^{\mu }-m_{l}\right)\psi_l,
\end{eqnarray}
which contains the contributions of baryons  ($n$ and $p$) and leptons  ($e$ and $\mu$).
$W^{\mu\nu}$ and $R^{a\mu\nu}$ are the antisymmetric field
tensors for $\omega^{\mu}$ and $\rho^{a\mu}$, respectively.
The parameters in the Lagrangian are usually determined by fitting nuclear matter
saturation properties and/or ground-state properties of finite nuclei. 

\begin{table*}[htp]
	\caption{Parameter $a_{\rho}$ generated from the NL3 model
		for different slope $L$ at saturation density $n_0$ without changing other saturation properties.
		The original NL3 model has $L=118.5$ MeV.}
	\label{tab:nl3l}
	\begin{center}
		\setlength{\tabcolsep}{2.6mm}{
			\begin{tabular}{lcccccccccc}
\hline\hline
$L$ (MeV) & 30.0    & 40.0    & 50.0    & 60.0    & 70.0  & 80.0  & 90.0  & 100.0  & 110.0 & 118.5\\
\hline
$a_\rho$  & 0.7537 & 0.6686 & 0.5835 & 0.4983 & 0.4132  & 0.3280 & 0.2429 & 0.1578 & 0.0726 & 0\\
				\hline\hline
		\end{tabular}}
	\end{center}
\end{table*}

To study the effect of symmetry energy slope $L$, 
we generate a series of parameter sets with a density-dependent
isovector coupling based on the NL3 parametrization (referred to as NL3L).
The parameters and saturation properties of the original NL3 model are listed in 
Table~\ref{tab:nl3} and Table~\ref{tab:sat}, respectively.
In NL3L, the symmetry energy
slope parameter is tuned to be $L=30-110~\MeV$
at saturation density, as listed in Table~\ref{tab:nl3l}.
The NL3L parameter sets have the same saturation properties 
as the origin NL3 except with different $L$. 
The isovector coupling $g_\rho$ in NL3L is taken to be density dependent
as in the DDRMF approach,
\begin{eqnarray}
g_{\rho}(n_b)=g_{\rho}(n_0)\exp\left[-a_{\rho}\left(\frac{n_b}{n_0}-1\right)\right],
\label{eq:g_r}
\end{eqnarray}%
where $n_0$ is the saturation density. 
Through this way, the symmetry energy slope $L$ can be tailored conveniently 
by adjusting $a_{\rho}$ without affecting other saturation properties 
and leaving other parameters the same as the original ones.
The density dependence of $g_{\rho}$ contributes a rearrangement item for nucleons,
\begin{eqnarray}
\Sigma_{r}=\frac{1}{2}\sum_{i=p,n}\frac{\partial{g_{\rho}(n_b)}}{\partial{n_b}}\tau_3{n_i}{\rho}
=-\frac{1}{2}a_{\rho}g_{\rho}(n_b)\frac{n_p-n_n}{n_0}{\rho}.
\label{eq:er}
\end{eqnarray}%

In a homogeneous matter, the meson field equations have the following form:
\begin{eqnarray}
&&m_{\sigma }^{2}\sigma +g_{2}\sigma ^{2}+g_{3}\sigma
^{3}=-g_{\sigma }\left( n_{p}^{s}+n_{n}^{s}\right) ,
\label{eq:eqms} \\
&&m_{\omega }^{2}\omega 
=g_{\omega}\left( n_{p}+n_{n}\right) ,
\label{eq:eqmw} \\
&&m_{\rho }^{2}{\rho}
=\frac{g_{\rho }(n_b)}{2}\left(n_{p}-n_{n}\right) ,
\label{eq:eqmr}
\end{eqnarray}%
where $n_i^s$ and $n_i$ represent the scalar and vector densities
of the $i$th baryon ($i=n, p$), respectively.
The equations of motion for nucleons give the standard relations between the densities
and chemical potentials,
\begin{eqnarray}
\mu_{p} &=& {\sqrt{\left( k_{F}^{p}\right)^{2}+{M^{\ast }}^{2}}}+g_{\omega}\omega+\Sigma_{r}   +\frac{g_{\rho}(n_b)}{2}\rho,
\label{eq:mup} \\
\mu_{n} &=& {\sqrt{\left( k_{F}^{n}\right)^{2}+{M^{\ast }}^{2}}}+g_{\omega}\omega+\Sigma_{r}  -\frac{g_{\rho}(n_b)}{2}\rho,
\label{eq:mun}
\end{eqnarray}%
where $M^{\ast}=M+g_{\sigma}\sigma$ is the effective nucleon mass,
and $k_{F}^{i}$ is the Fermi momentum of species $i$, which is related to the vector density 
by $n_i=\left(k_{F}^{i}\right)^3/3\pi^2$. For neutron-star matter in $\beta$ equilibrium,
the chemical potentials satisfy the relations $\mu_{p}=\mu_{n}-\mu_{e}$ and $\mu_{\mu}=\mu_{e}$,
where the chemical potentials of leptons are given by $\mu_{l}=\sqrt{\left({k_{F}^{l}}\right)^{2}+m_{l}^{2}}$.
In neutron-star matter, the total energy density and pressure are given by
\begin{eqnarray}
\varepsilon &=&\sum_{i=p,n}\frac{1}{\pi^2}
     \int_{0}^{k^{i}_{F}}{\sqrt{k^2+{M^{\ast}}^2}}k^2dk   \nonumber \\
&& + \frac{1}{2}m^2_{\sigma}{\sigma}^2+\frac{1}{3}{g_2}{\sigma}^3
     +\frac{1}{4}{g_3}{\sigma}^4
 + \frac{1}{2}m^2_{\omega}{\omega}^2 \nonumber  \\
&&   + \frac{1}{2}m^2_{\rho}{\rho}^2
         + \varepsilon_l,   \\
     \label{eq:ehp}
P &=& \sum_{i=p,n}\frac{1}{3\pi^2}\int_{0}^{k^{i}_{F}}
      \frac{1}{\sqrt{k^2+{M^{\ast}}^2}}k^4dk    \nonumber  \\
&& - \frac{1}{2}m^2_{\sigma}{\sigma}^2-\frac{1}{3}{g_2}{\sigma}^3
     -\frac{1}{4}{g_3}{\sigma}^4      + \frac{1}{2}m^2_{\omega}{\omega}^2
         \nonumber \\
      && + \frac{1}{2}m^2_{\rho}{\rho}^2
      +n_b{\Sigma_{r}}+P_l,
           \label{eq:php}
\end{eqnarray}
where $\varepsilon_l$ and $P_l$ ($l=e, \mu$) are the energy density and the pressure
from leptons, respectively. With given baryon number density $n_b$, the EOS can be derived by solving the meson field equations under the conditions of $\beta$ equilibrium and charge neutrality.

\begin{figure}[htp]
\begin{minipage}[t]{1\linewidth}
\begin{center}
\includegraphics[bb=5 5 580 480, width=8 cm,clip]{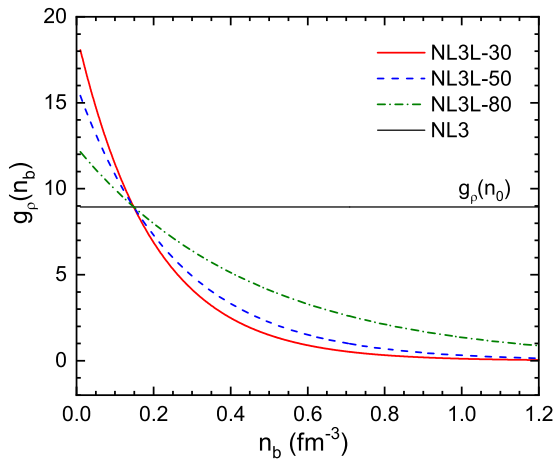}
\end{center}
\caption{Isovector coupling $g_\rho(n_b)$ as a function of the baryon number density $n_b$.}
\label{fig:1nbgr}
  \end{minipage}
\end{figure}

\section{The results and discussion}
\label{sec:results}
In this section, we investigate the symmetry energy effect on EOS and the properties of neutron star. 
\begin{figure*}[htp]
	\includegraphics[bb=40 5 580 500, width=8 cm,clip]{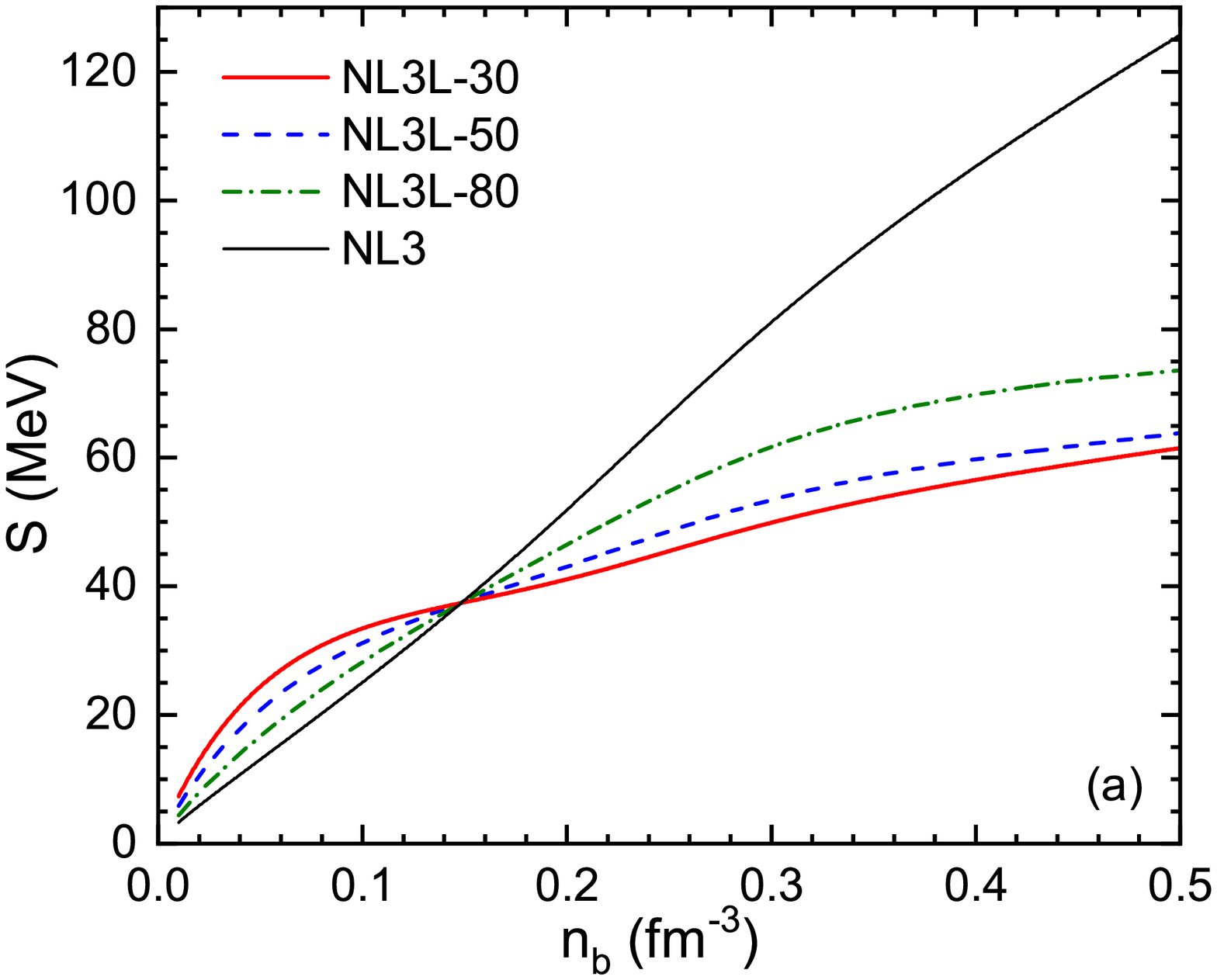}
	\includegraphics[bb=40 5 580 500, width=8 cm,clip]{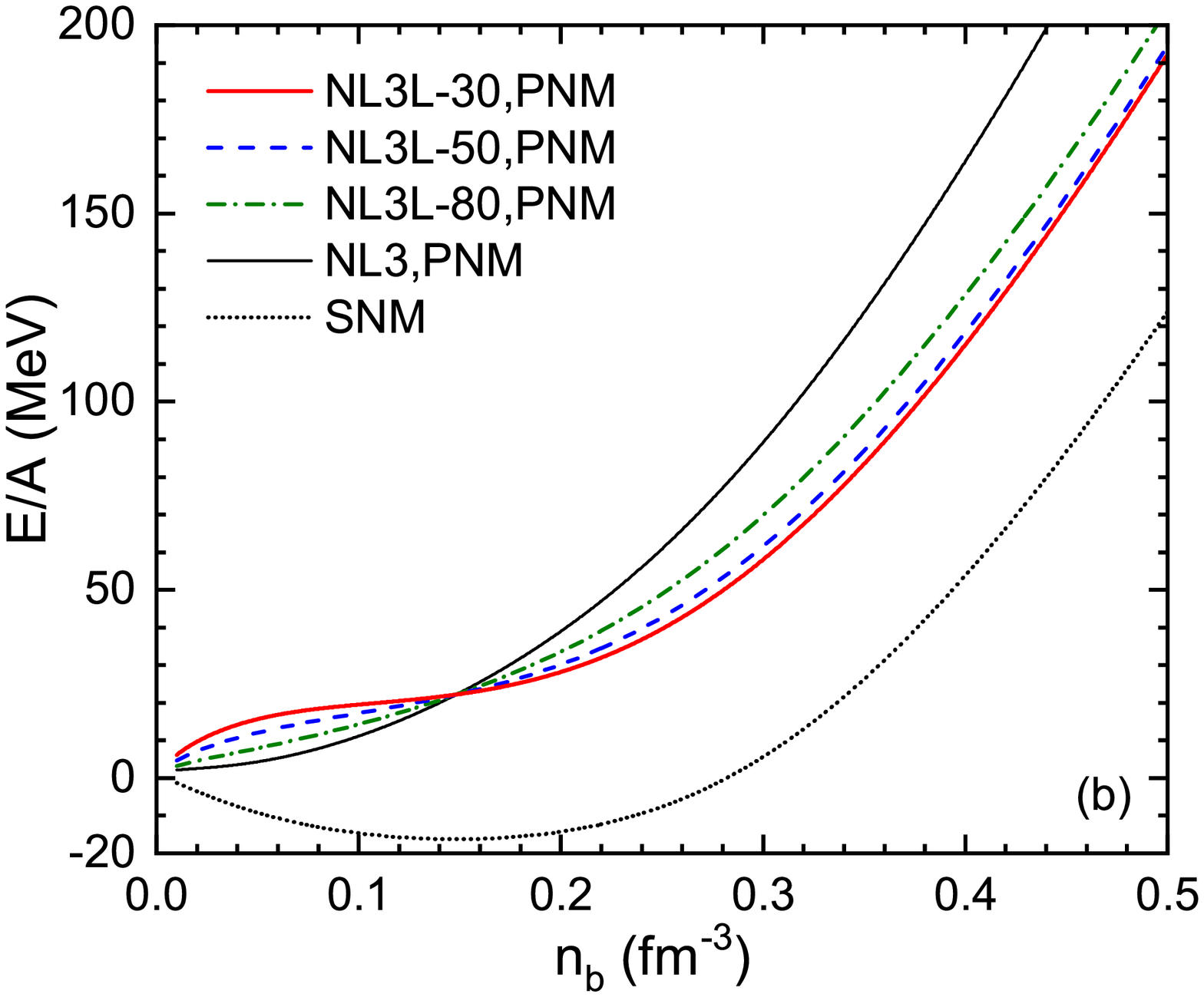}
	\caption{Symmetry energy $S$ (a) and energy per nucleon $E/A$ in pure neutron matter (PNM) and symmetry nuclear matter (SNM) (b) as functions of the baryon number density.}
	\label{fig:2nbs}
\end{figure*}
It is well known that the symmetry energy slope $L$ can significantly affect the neutron-star radius and tidal deformability.
However, its value is still very uncertain and cannot be well constrained from current observations~\cite{Oertel2017}.
We apply the RMFL model to generate NL3L parameter sets with different values of $L$.
In Fig.~\ref{fig:1nbgr}, we plot the density-dependent behavior of the isovector coupling $g_{\rho}(n_b)$ 
as a function of the baryon number density $n_b$.
The original NL3 parameter set has fixed $g_\rho(n_b)=g_\rho(n_0)$. 
In the following discussions, we investigate the symmetry energy effect by comparing the results of $L=30, 50, 80$~MeV (named as NL3L-30, NL3L-50, and NL3L-80) 
and the original NL3 parametrization which has $L=118.5$~MeV.
It can be found that $g_{\rho}(n_b)$ decreases as the density $n_b$ increases, and a smaller $g_\rho$ results in a smaller symmetry energy.
At subnuclear densities ($n_b < n_0$), the coupling constant $g_{\rho}(n_b)$ is bigger than $g_{\rho}(n_0)$, 
while it becomes smaller at high densities.
This trend can be easily understood from Eq.~(\ref{eq:g_r}).
Furthermore, a smaller $L$ corresponds to more rapid decrease of $g_\rho$.

\begin{figure*}[htp]
	\includegraphics[bb=40 5 580 500, width=8 cm,clip]{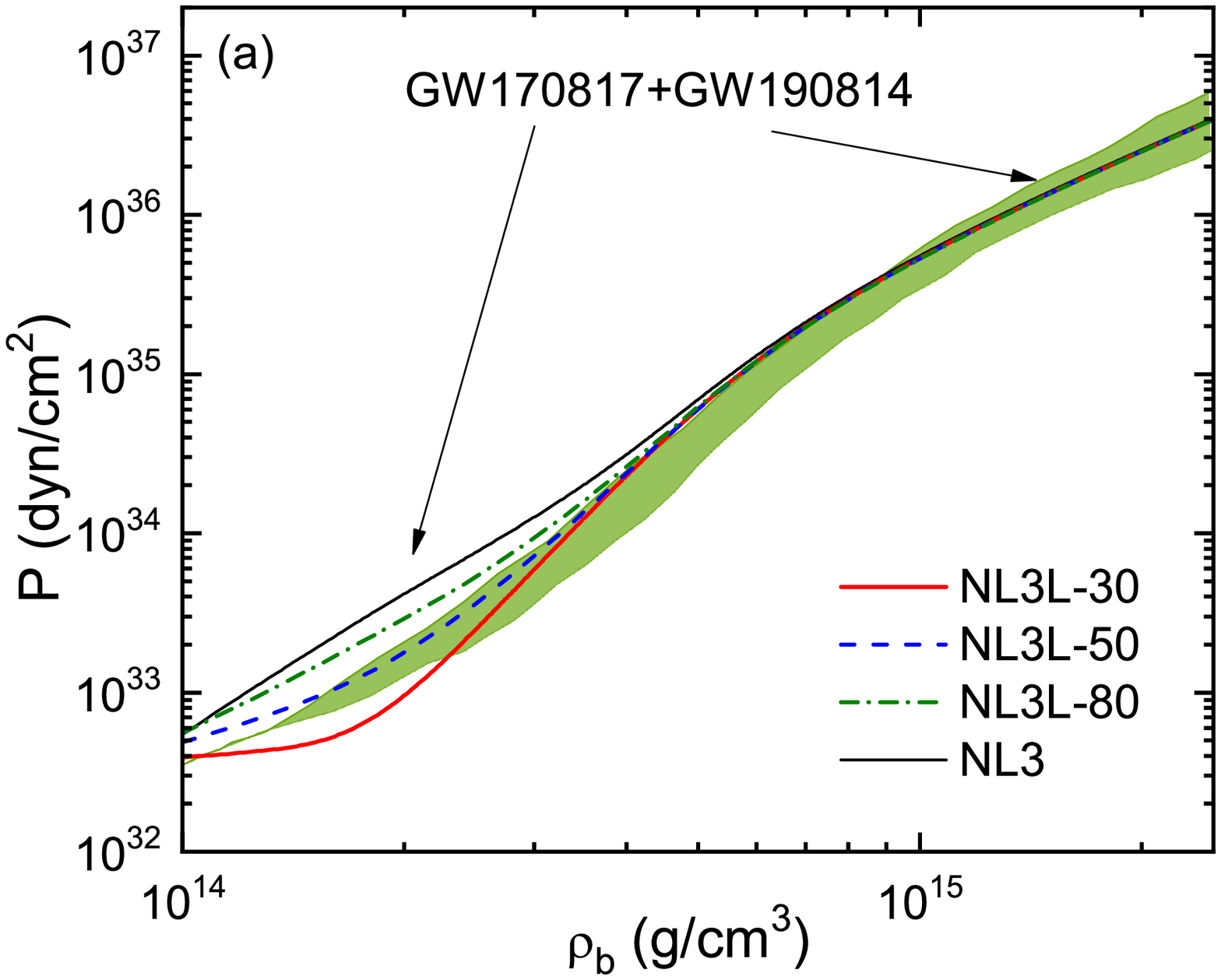}%
	\includegraphics[bb=40 5 580 500, width=8 cm,clip]{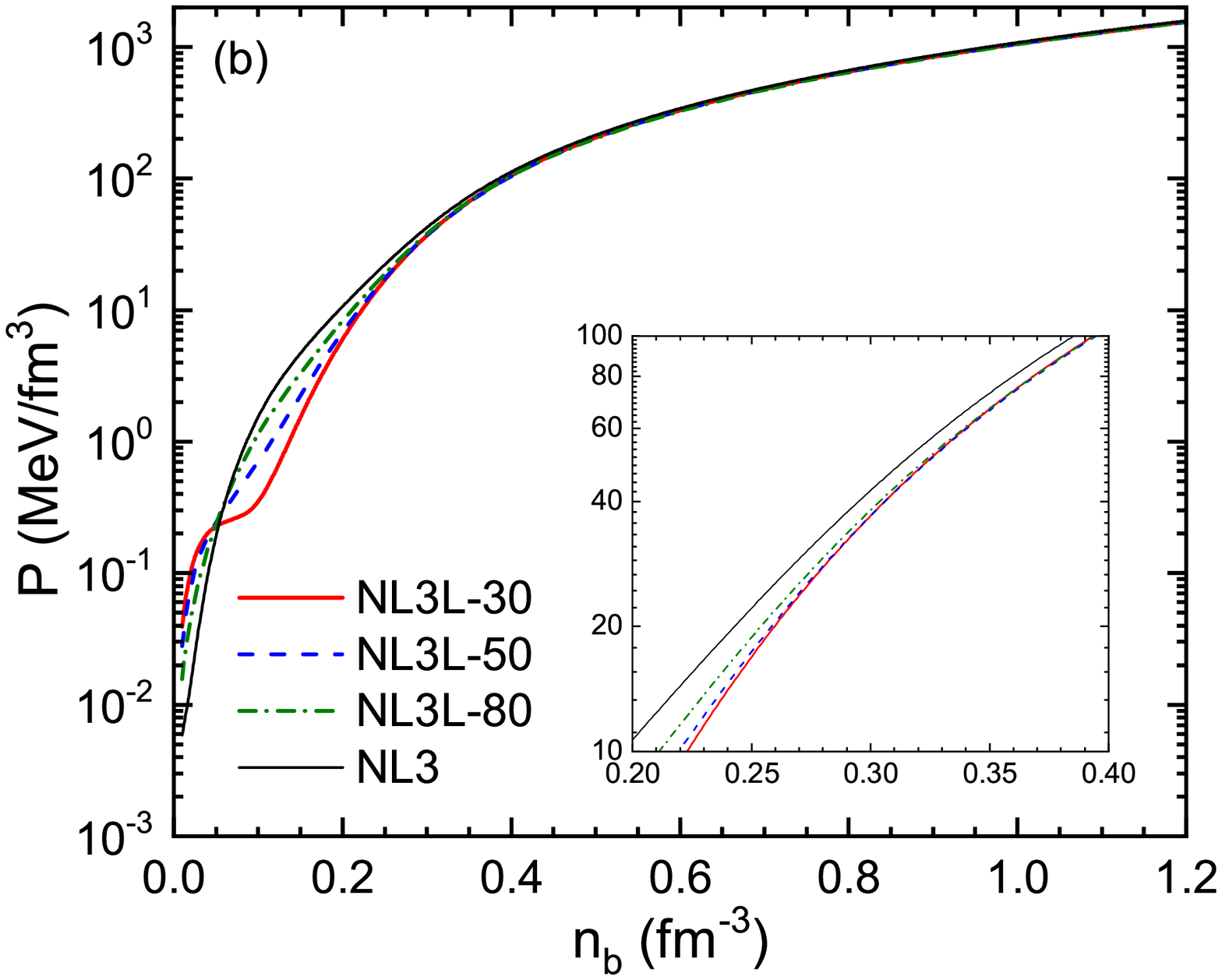}
	\caption{
		Pressures as functions of the baryon mass density $\rho_b$ (a) and number density $n_b$ (b) obtained using different parameter sets.}
	\label{fig:3nbp}
\end{figure*}
We plot the symmetry energy $S$ and the energy per nucleon $E/A$ as functions of baryon number density in the left and right panels of Fig.~\ref{fig:2nbs} . 
The results of NL3L-30, NL3L-50, NL3L-80 and the original NL3 parameter sets are shown. 
It is seen that the symmetry energy with a smaller $L$ is lower (higher) than that with a larger $L$ at
$n_b > n_0$ ($n_b < n_0$), similar behavior is observed in Ref.~\cite{Bao2014}.
Different NL3L parametrizations have the same energy ($E/A$) for symmetry nuclear matter (SNM), 
since the isovector interaction vanishes in SNM (see Eq.~\ref{eq:eqmr}). 
In contrast, $E/A$ in pure neutron matter (PNM) shows a significant dependence on $L$.
The $L$-dependence of $E/A$ in PNM is very similar to that of $S$, which can be understood from
the relation $E/A (\textrm{PNM}) \approx E/A (\textrm{SNM}) + S$.

\begin{figure}[htp]
	\includegraphics[bb=20 5 600 500, width=8 cm,clip]{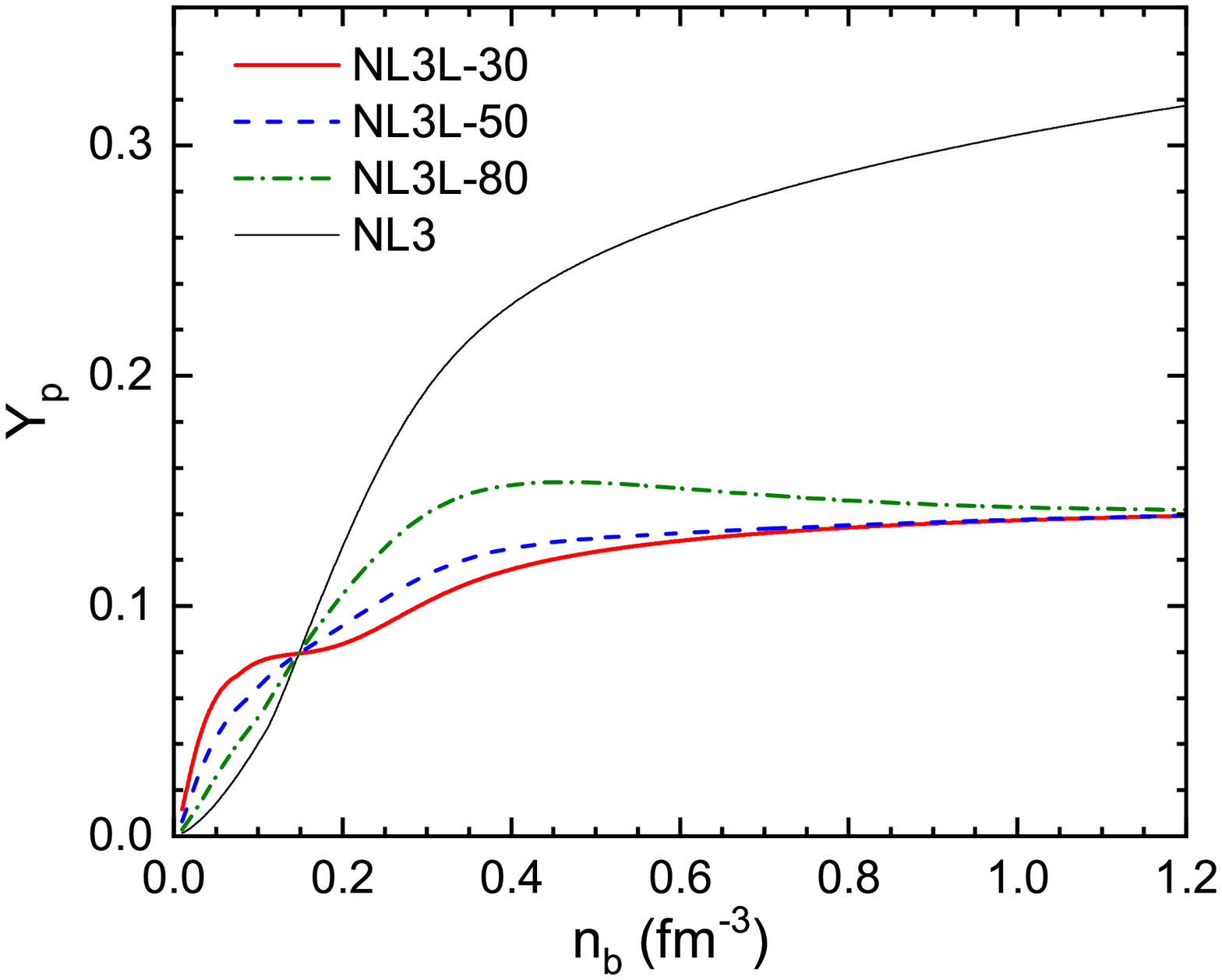}%
	\caption{Proton fraction $Y_p$ as a function of baryon number density.}
	\label{fig:4nbyp}
\end{figure}

In Fig.~\ref{fig:3nbp}, we show the EOSs with different $L$ as functions of 
the baryon mass density $\rho_b$ (left panel) and number density $n_b$ (right panel), respectively.
The green shaded area indicates the constraint by assuming the secondary component of GW190814 is a neutron star~\cite{Abbott2020b}.
It is seen that the results of NL3L-30 and NL3L-50 are more consistent with the constraint than the original NL3 parametrization.
From the right panel, we see that there are visible differences
among different $L$ lines at low densities ($n_b < 0.3$ fm$^{-3}$), 
while all NL3L and NL3 parameter sets result in very similar EOSs at high densities.
This behavior is related to
the density dependence of $g_\rho$ shown in Fig.~\ref{fig:1nbgr}. 
At low densities, the differences in $g_\rho$ correspond to different isovector contributions, which yield different EOSs.
However, $g_\rho$ of all NL3L parametrizations decrease with increasing density and approach zero at high densities, so the differences caused by $\rho$ meson tend to disappear.
Considering the constraints shown in the left panel, it is clear that a stiff enough EOS like NL3L is helpful
to support massive neutron star and a proper small $L$ is favored at low density. 
To better understand the high-density EOS behavior, 
we show the proton fraction $Y_p$ under $\beta$ equilibrium as a function of baryon number density in Fig.~\ref{fig:4nbyp}. 
One can see that $Y_p$ obtained in NL3L parametrizations with various $L$ are different from each other at low densities, 
but they become very close at high densities. 
However, $Y_p$ in the original NL3 model is obviously different from those of NL3L, 
since $g_\rho$ has different behavior as shown in Fig.~\ref{fig:1nbgr}.
The pressure is mainly contributed by nuclear Fermi energy and  interaction.
On one hand, at high densities, the contribution of isovector interaction to pressure decreases. On the other hand, the $Y_p$ obtained by different $L$ tend to be the same, which lead to close proton and neutron Fermi energies. Hence the  pressure from NL3L are indistinguishable at high densities.  The original NL3 model has different trends of $Y_p$ and $g_\rho$ from NL3L, which leads to visible difference of pressure (see the insert of Fig.~\ref{fig:3nbp} (b)).

\begin{figure*}[htp]
	\includegraphics[bb=20 5 600 580, width=8 cm,clip]{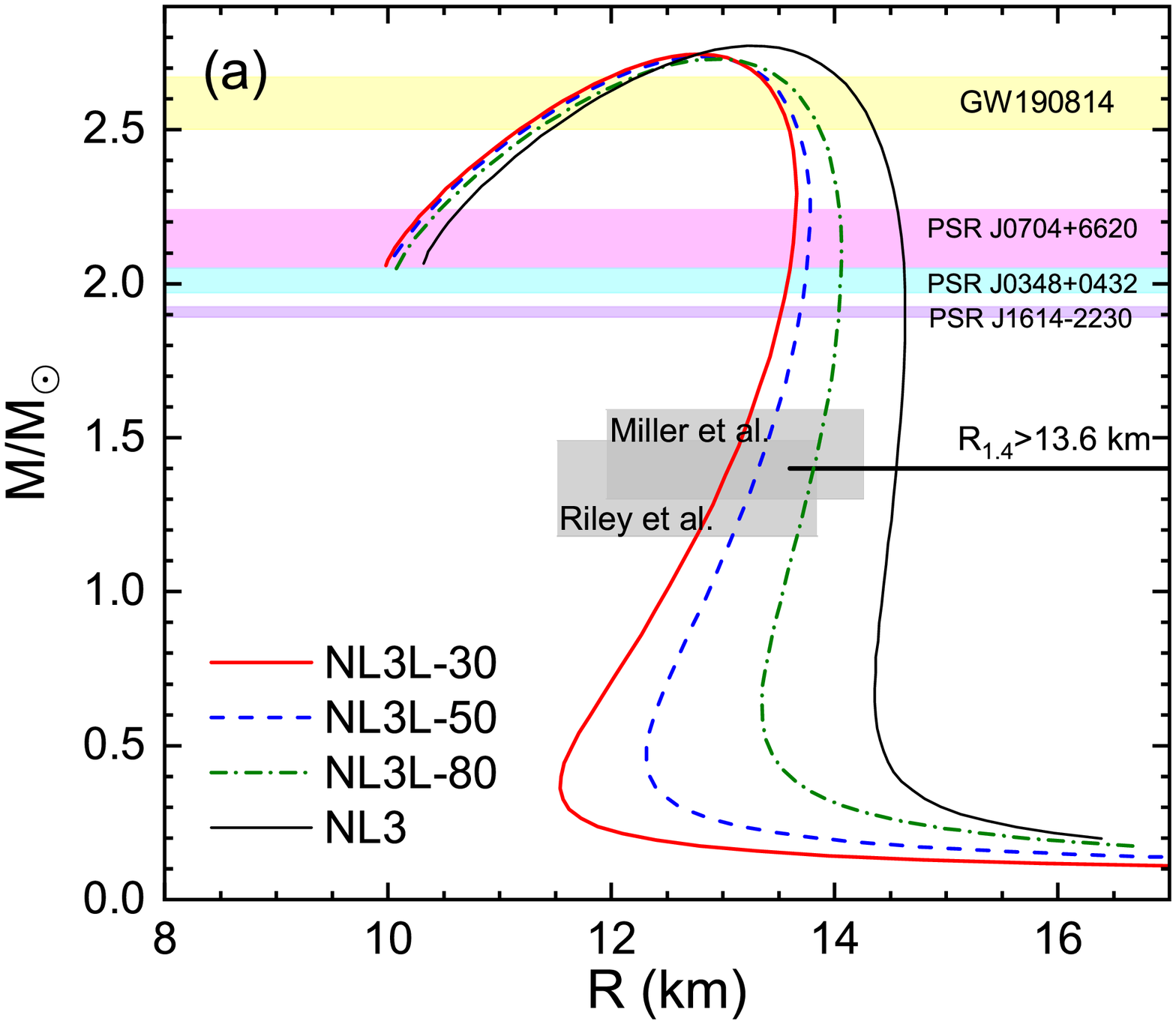}%
	\includegraphics[bb=20 5 600 580, width=8 cm,clip]{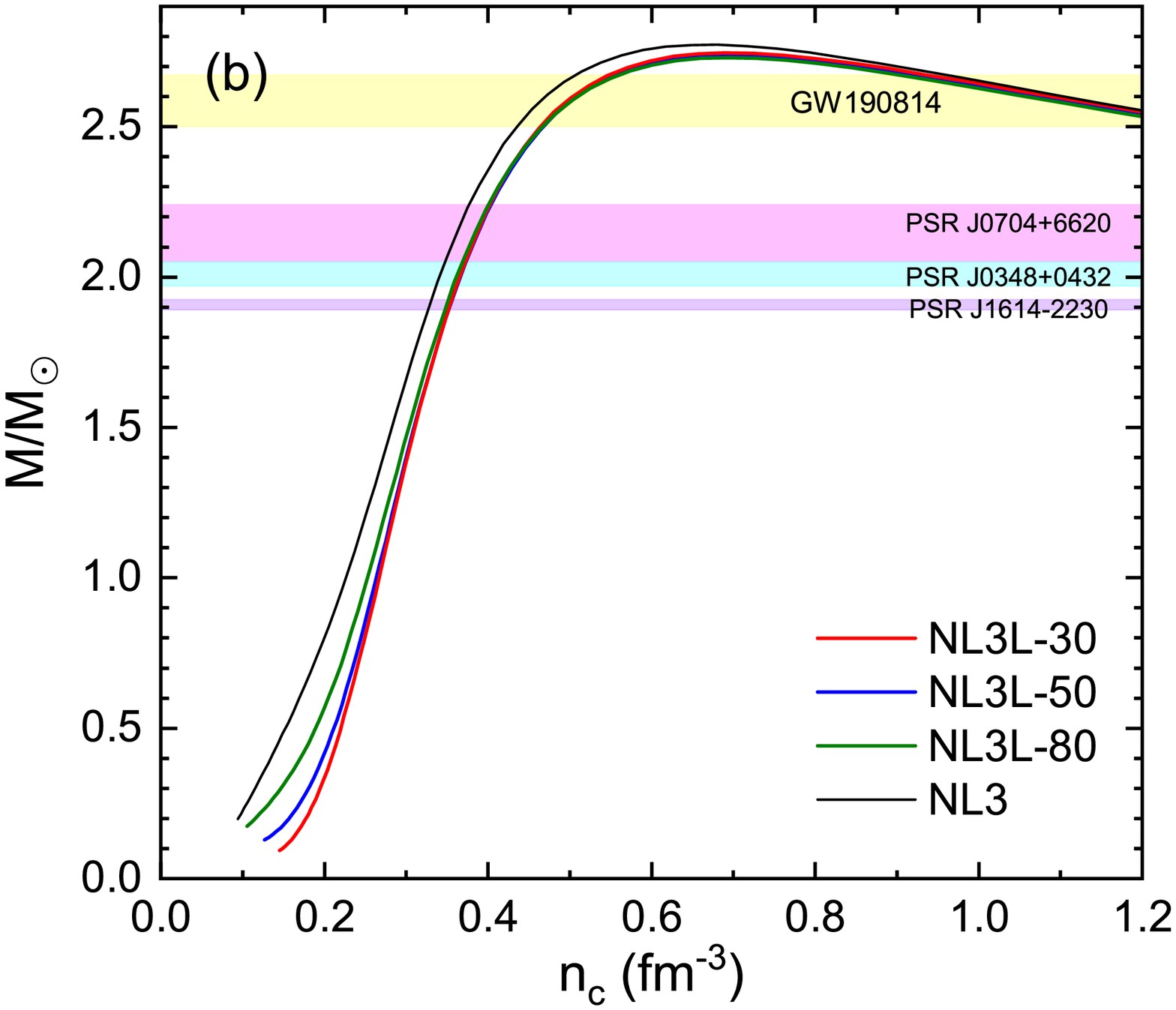}
	\caption{ Neutron-star masses as functions of the radius $R$ (a) and the central density $n_c$ (b). 
		Constraints from astronomical observations are also shown.}
	\label{fig:5mr}
\end{figure*}
\begin{figure}[htp]
	\includegraphics[bb=20 5 580 580, width=8 cm,clip]{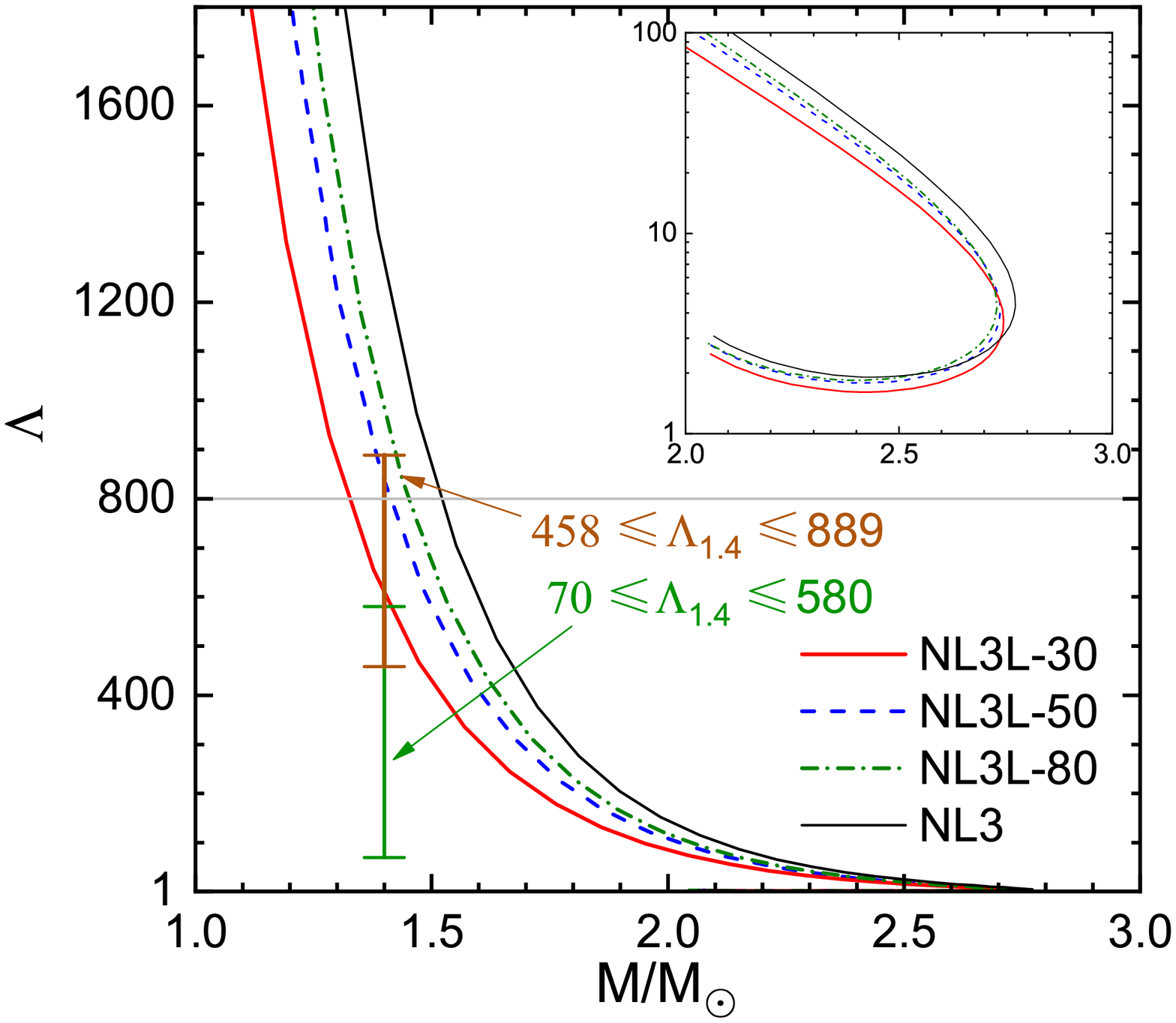}
	\caption{Tidal deformability as a function of the neutron-star mass for different parameter sets. 
		Constraints on $\Lambda_{1.4}$ from astronomical observations are also shown.}
	\label{fig:6ml}
\end{figure}

To examine the effect of symmetry energy slope $L$ on the properties of neutron stars, 
we solve the Toman-Oppenheimer-Volkoff (TOV) equations by using the NL3L parametrizations and the original NL3 model. 
The observed PSR J1614-2230, PSR J0348+0432, and PSR J0704+6620 suggest the neutron-star maximum mass $M_\textrm{TOV}$ should be at least larger than 2~$M_\odot$.
Using the EOS of neutron-star matter, the original NL3 parameter set predicts a maximum mass of 2.77~$M_\odot$. 
In Fig.~\ref{fig:5mr}, the mass-radius relation with different $L$ is presented in the left panel (a), 
and the neutron-star mass as a function of the neutron-star central density $n_c$ is shown in the right panel (b). 
Within the NL3 and NL3L models, the neutron-star maximum mass could be larger than 2.6~$M_\odot$,
though with $L=30, 50, 80$~MeV, there exists a tiny decrease of $M_\textrm{TOV}$ (Table~\ref{tab:ns-properties}). 
The NL3L-30 and NL3L-50 parameter sets could satisfy the constraints from NICER and the estimation of $R_{1.4} \le 13.6$~km from GW170817. 
It is shown that smaller $L$ corresponds to smaller radius, 
but this $L$-dependence becomes much weaker for massive neutron stars.
Besides, the $L$ effect on the relation of the neutron-star central density $n_c$ and mass is also small except at lower central densities. 
This is because $L$ affects the EOSs at the low-density region $n_b \le 0.3$~ fm$^{-3}$ while EOSs keep almost the same at high density.
The maximum mass neutron star has central density of $n_c \sim 0.7$~fm$^{-3}$.
The tidal deformability as a function of neutron-star mass is plotted in Fig.~\ref{fig:6ml}, 
and the details of massive neutron stars are shown in logarithmic coordinates in the upper right corner. 
The orange-colored constraint $458 \le \Lambda_{1.4} \le 889$ in the figure comes from a presupposition that the secondary component of GW190814 is a neutron star. 
It is shown that the results of NL3L-30 and NL3L-50 meet this limit. 
The results of $L \ge 50$~MeV are beyond the 
$\Lambda_{1.4} \le 800$ constraint. 
However, even NL3L-30 has $\Lambda_{1.4}=608.8$ also lies outside the advanced constraint $70 \le \Lambda_{1.4} \le 580$ (green colored in the figure). 

\begin{table}[htp]
\caption{Neutron star properties predicted by the NL3 and NL3L models}
\begin{center}
\setlength{\tabcolsep}{2.1 mm}{
\begin{tabular}{lcccccccccccc}
\hline\hline
Model   &NL3L-30  &NL3L-50  &NL3L-80  &NL3  \\
\hline
$M_\textrm{TOV}$ ($M_\odot$)     &2.75  &2.74  &2.73  &2.77  \\
\hline
$R_{1.4}$ (km)           &13.04  &13.31  &13.80  &14.55   \\
\hline
$\Lambda_{1.4}$   &608.8  &815.9  &975.2  &1242.4  \\
\hline
$n_c$ (fm$^{-3}$)           & 0.689 & 0.690 &  0.693 &  0.680  \\
\hline\hline
\end{tabular}}
\label{tab:ns-properties}
\end{center}
\end{table}


\section{Summary}
\label{sec:summary}

In summary, motivated by the observation of the 2.6~$M_\odot$ object of GW190814,
we have explored if such a massive object could be a neutron star. 
Since the lack of an electromagnetic counterpart to GW190814 and no tidal distortions observed from gravitational waveform, 
the secondary component of GW190814 may be a black hole or the heaviest neutron star observed until now.
We apply the RMFL model with the NL3L parametrizations to describe the neutron-star matter. 
To study the effect of symmetry energy and its slope, we employ a
density-dependent coupling $g_\rho(n_b)$ as in the DDRMF approach. 

Using a group of NL3L parameter sets, we have investigated 
the symmetry energy effect on the EOS and the properties of neutron stars. 
It was found that a smaller $L$ (NL3L-30, NL3L-50) corresponds to smaller symmetry energy at $n_b > n_0$,
which implies lower pressure in neutron-star matter. 
However, at sufficiently high density the $L$-dependence of EOSs tends to disappear due to rather small values of $g_\rho$.
We found that these NL3L parametrizations provide EOSs that are stiff enough 
to support $M_\textrm{TOV}>2.6~M_\odot$. 
The resulting EOSs of NL3L-30 and NL3L-50 lie roughly 
in the area where the secondary component of GW190814 is assumed to be a neutron star.
The neutron-star radii using NL3L-30, NL3L-50 and NL3L-80 are consistent with NICER constraints, while the results of NL3L-30 and NL3L-50 meet
the constraints of $\Lambda_{1.4}=616_{-158}^{+273}$ and $R_{1.4}=12.9_{-0.7}^{+0.8}$~km that corresponds to GW190814 neutron star-black hole assumption.
However, even with the smallest $L$=30~MeV (NL3L-30), we obtained $\Lambda_{1.4}=608.8$ 
that is slightly larger than the constraint $70 \le \Lambda_{1.4} \le 580$ from GW170817.
Since the maximum mass $M_\textrm{TOV}$ is mainly determined by the high-density EOS, 
all NL3L parametrizations can support 2.6~$M_\odot$ neutron star.
Meanwhile, a smaller $L$ results in smaller radii of neutron stars, which can be compatible with current observations.
Based on the present work, we cannot rule out the possibility that the secondary object of GW190814 is a neutron star.
More precise measurement of tidal deformability by the gravitational wave detectors may help to constrain the EOS of neutron-star matter in the future.

\section*{Acknowledgment}
This work is supported by the National Key R$\&$D Program of China (Grant Nos. 2018YFA0404703, and 2017YFA0402602), 
and the National Natural Science Foundation of China
(Grants No. 11673002, No. U1531243, and No. 11805115).


\newpage


\begin{thebibliography}{99}
\bibitem{Abbott2016}  B. P. Abbott et al. (LIGO Scientific Collaboration and
Virgo Collaboration), 
Phys. Rev. Lett. 116, 061102 (2016).
\bibitem{Abbott2017} B. P. Abbott et al. (LIGO Scientific Collaboration and Virgo
Collaboration), 
Phys. Rev. Lett. 119, 161101 (2017).
\bibitem{Abbott2020a} B. P. Abbott et al. (LIGO Scientific and Virgo Collaborations), 
Astrophys. J. Lett. 900, L13 (2020).
%
\bibitem{Abbott2020b} R. Abbott et al. (LIGO Scientific and Virgo Collaborations), 
Astrophys. J. Lett. 896, L44 (2020).
%
\bibitem{Cromartie2020} H. T. Cromartie et al., 
Nat. Astron. 4, 72 (2020).
%

\bibitem{Wu2020} X. H. Wu, S. Du and R. X. Xu,
MNRAS 499, 4526–4533 (2020).
%
\bibitem{Godzieba2020} D. A. Godzieba, D. Radice and S. Bernuzzi, 
arXiv:2007.10999
%
\bibitem{Fattoyev2020} F. J. Fattoyev, C. Horowitz, J. Piekarewicz, B. Reed,  
Phys. Rev. C 102, 065805 (2020).
%
\bibitem{Tan2020} H. Tan, J. Noronha-Hostler, N. Yunes, 
Phys. Rev. Lett. 125, 261104 (2020).
%
\bibitem{McLerran2019} L. McLerran and S. Reddy, 
Phys. Rev. Lett. 122, 122701 (2019).
%
\bibitem{Fattoyev2018} F. J. Fattoyev, J. Piekarewicz, C. J. Horowitz,
Phys. Rev. Lett. 120, 172702 (2018).
\bibitem{Hu2020}
J. N. Hu, S. S. Bao, Y. Zhang, K. Nakazato, K. Sumiyoshi, H. Shen,
Prog. Theor. Exp. Phys., 4, 043D01 (2020).
%
\bibitem{Centelles2009} M. Centelles, X. Roca-Maza, X. Vinas, M. Warda,
Phys. Rev. Lett. 102, 122502 (2009).
%
\bibitem{Steiner2012} A. W. Steiner, S. Gandolfi, 
Phys. Rev. Lett. 108, 081102 (2012).
\bibitem{Lattimer2013} J. M. Lattimer, Y. Lim, 
Astrophys. J. 771, 51 (2013).
\bibitem{Tews2017} I. Tews, J. M. Lattimer, A. Ohnishi, and E. E. Kolomeitsev,
Astrophys. J. 848, 105 (2017).
\bibitem{Oertel2017} M. Oertel,  M. Hempel, T. Kl\"{a}hn,  S. Typel, 
  Rev. Mod. Phys. 89, 015007 (2017).

%
\bibitem{Bao2014} S. S. Bao, J. N. Hu, Z. W. Zhang, and H. Shen, 
Phys. Rev. C 90, 045802 (2014).
%
\bibitem{Drago2014} A. Drago, A. Lavagno, and G. Pagliara, 
Phys. Rev. D 89, 043014 (2014).
%

\bibitem{Spinella2017}  W. M. Spinella, 
A Systematic Investigation of Exotic Matter in Neutron Stars, Ph.D. thesis, Claremont Graduate University \& San Diego State University (2017).
%
\bibitem{Demorest2010} P. B. Demorest, T. Pennucci, S. M. Ranson, M. S. E.
Roberts, and J. W. T. Hessels, Nature (London) 467, 1081 (2010).
\bibitem{Fonseca2016} E. Fonseca et al., 
Astrophys. J. 832, 167 (2016).
\bibitem{Arzoumanian2018} Z. Arzoumanian et al., 
Astrophys. J. Suppl. 235, 37 (2018).
%
\bibitem{Antoniadis2013} J. Antoniadis,  P. C. C. Freire,  N. Wex, et al., 
Science, 340, 448 (2013).
%
\bibitem{Linares2018} M. Linares, T. Shahbaz, J. Casares, 
Astrophys. J. 859, 54 (2018).
%
\bibitem{Abbott2018} B. P. Abbott et al. (LIGO Scientific Collaboration and
Virgo Collaboration), 
Phys. Rev. Lett. 121, 161101 (2018).
%
\bibitem{Annala2018}  E. Annala, T. Gorda, A. Kurkela, A. Vuorinen, 
Phys. Rev. Lett. 120, 172703 (2018).
\bibitem{Most2018} E. R. Most, L. R. Weih, L. Rezzolla, and J. Schaffner-Bielich, 
Phys. Rev. Lett. 120, 261103 (2018).
%
\bibitem{Raaijmakers2019} G. Raaijmakers et al.,
Astrophys. J. Lett. 887, L22 (2019).
%
\bibitem{Miller2019} M. C. Miller et al.,
Astrophys. J. Lett. 887, L24 (2019).
%
\bibitem{Riley2019} T. E. Riley et al.,
Astrophys. J. Lett. 887, L21 (2019).
%
\bibitem{Margalit2017} B. Margalit, B.D. Metzger, 
Astrophys. J. Lett. 850, L19 (2017).
\bibitem{Rezzolla2018} L. Rezzolla, E.R. Most, L.R. Weih, 
Astrophys. J. Lett. 852, L25 (2018).
\bibitem{Ruiz2018} M. Ruiz, S.L. Shapiro, A. Tsokaros, 
Phys. Rev. D 97, 021501 (2018).
\bibitem{Shibata2019} M. Shibata, E. Zhou, K. Kiuchi, S. Fujibayashi,
Phys. Rev. D 100, 023015 (2019).
\bibitem{Abbott2020c}  B. P. Abbott et al. (LIGO Scientific Collaboration and
Virgo Collaboration), 
Classical and Quantum Gravity, Vol. 37, No 4, p 045006 (2020).
%
\bibitem{Lalazissis1997} G. A. Lalazissis, J. K\"{o}nig and P. Ring, 
Phys. Rev. C 55, 540  (1997).
%
\bibitem{Wu2018} X. H. Wu, A. Ohnishi, H. Shen,
Phys. Rev. C 98, 065801 (2018).
\bibitem{Ishizuka2008} C. Ishizuka, A. Ohnishi, K. Tsubakihara, K. Sumiyoshi and S. Yamada, 
J. Phys. G 35, 085201 (2008).
%
\bibitem{Wu2017} X. H. Wu and H. Shen,
Phys. Rev. C 96, 025802 (2017).


\bibliographystyle{unsrt}
\end{thebibliography}
\end{document}